\documentclass{elsart}

\usepackage{graphicx}
\usepackage{amssymb}

\newcommand{\ud}{\mathrm{d}}
\def\NIMA#1#2#3{Nucl. Instr. Meth. Phys. Res. A {\bf #1}\ (#2)\ #3}
\def\NIM#1#2#3{Nucl. Instr. Meth. {\bf #1}\ (#2)\ #3}
\def\IEEE#1#2#3{IEEE Trans. Nucl. Sci. vol. {\bf #1}\ (#2)\ #3}
\def\ARNPS#1#2#3{Ann. Rev. Nucl. Part. Sc. {\bf #1}\ (#2)\ #3}

\begin{document}


\begin{frontmatter}

\title{Energy loss of pions and electrons of 1 to 6 GeV/c in drift chambers
operated with Xe,CO$_2$(15\%)}

\author[gsi]{A.~Andronic\thanksref{info}\thanksref{leave}},
\author[hei]{H.~Appelsh{\"a}user}, 
\author[gsi]{C.~Blume}, 
\author[gsi]{P.~Braun-Munzinger}, 
\author[mue]{D.~Bucher}, 
\author[gsi]{O.~Busch}, 
\author[buc,hei]{V.~C{\u a}t{\u a}nescu}, 
\author[buc,gsi]{M.~Ciobanu}, 
\author[gsi]{H.~Daues}, 
\author[hei]{D.~Emschermann}, 
\author[dub]{O.~Fateev},
\author[gsi]{Y.~Foka}, 
\author[gsi]{C.~Garabatos}, 
\author[tok]{T.~Gunji}, 
\author[hei]{N.~Herrmann}, 
\author[tok]{M.~Inuzuka}, 
\author[dub]{E.~Kislov},
\author[kip]{V.~Lindenstruth},
\author[hei]{W.~Ludolphs}, 
\author[hei]{T.~Mahmoud}, 
\author[hei]{V.~Petracek}, 
\author[buc]{M.~Petrovici},
\author[hei]{I.~Rusanov}, 
\author[gsi]{A.~Sandoval},
\author[mue]{R.~Santo},
\author[hei]{R.~Schicker},  
\author[gsi]{R.S.~Simon}, 
\author[dub]{L.~Smykov},
\author[hei]{H.K.~Soltveit}, 
\author[hei]{J.~Stachel}, 
\author[gsi]{H.~Stelzer}, 
\author[gsi]{G.~Tsiledakis}, 
\author[hei]{B.~Vulpescu}, 
\author[mue]{J.P.~Wessels}, 
\author[hei]{B.~Windelband},
\author[hei]{C.~Xu},
\author[mue]{O.~Zaudtke},
\author[dub]{Yu.~Zanevsky},
\author[dub]{V.~Yurevich}

\address[gsi]{Gesellschaft f{\"u}r Schwerionenforschung, Darmstadt, Germany}
\address[hei]{Physikaliches Institut der Universit{\"a}t Heidelberg, Germany}  
\address[mue]{Institut f{\"u}r Kernphysik, Universit{\"a}t M{\"u}nster, Germany}
\address[buc]{NIPNE Bucharest, Romania}  
\address[dub]{JINR Dubna, Russia}  
\address[tok]{University of  Tokyo, Japan}  
\address[kip]{Kirchhoff-Institut f\"ur Physik, Heidelberg, Germany}

{for the ALICE Collaboration}

\thanks[info]{Corresponding author: GSI, Planckstr. 1, 64291 Darmstadt,
Germany; Email:~A.Andronic@gsi.de; Phone: +49 615971 2769; 
Fax: +49 615971 2989.}
\thanks[leave]{On leave from NIPNE Bucharest, Romania. }

\begin{abstract}
We present measurements of the energy loss of pions and electrons 
in drift chambers operated with a Xe,CO$_2$(15\%) mixture.
The measurements are carried out for particle momenta from 1 to 6 GeV/c 
using prototype drift chambers for the ALICE TRD.
Microscopic calculations are performed using input parameters calculated
with GEANT3. These calculations reproduce well the measured average 
and most probable values for pions, but a higher Fermi plateau is 
required in order to reproduce our electron data.
The widths of the measured distributions are smaller for data compared
to the calculations.
The electron/pion identification performance using the energy loss
is also presented.
\end{abstract}


\end{frontmatter}

\section{Introduction} \label{d:intro}

The ALICE Transition Radiation Detector (TRD) \cite{aa:tdr} is designed
to provide electron identification and particle tracking in the
high-multiplicity heavy-ion collisions at the LHC.
To achieve the challenging goals of the detector, accurate pulse height 
measurement in drift chambers operated with Xe,CO$_2$(15\%) over the 
drift time of about 2~$\mu$s 
is a necessary requirement.
For such precision measurements, it is of particular importance first 
to collect \cite{aa:att} and then to properly amplify \cite{aa:gain}
all the charge deposited in the detector.
For electrons, the transition radiation (TR), produced in an especially
designed raditor, is overimposed on the ionization energy loss and helps 
crucially in improving the electron/pion separation.
A factor of 100 pion rejection for 90\% electron efficiency is the design 
goal of the detector and was demonstrated with prototypes \cite{aa:tdr}.
The measurements of ionization energy loss (dE/dx) in TRD will contribute 
to the identification of other charged particles, supplementing the 
identification power of the ALICE Time Projection Chamber.
A good understanding of dE/dx in the TRD is a prerequisite for high-precision
simulations of the detector performance in terms of TR.

Existing measurements of dE/dx in Xe-based mixtures 
\cite{aa:fis,aa:wal,aa:her} are scarce and have good precision on an absolute 
energy scale only in one case \cite{aa:her}.
Calculations of dE/dx can reproduce the measured data, in particular for 
Xe-based mixtures \cite{aa:her,aa:apo}.
Ref.~\cite{aa:ali} is an excellent early overview on energy loss 
measurements and calculations as well as on its application to particle 
identification. For a more recent account, see ref. \cite{aa:vav}.

We report on dE/dx measurements performed during prototype tests 
for the ALICE TRD. 
The experimental setup and method of data analysis are described in the 
next section. 
We then present the basic ingredients of our simulation code and discuss
the general outcome. 
The measured data in comparison to the calculations are presented in
Section~\ref{d:res}.

\section{Experimental setup} \label{d:meth} 

The results are obtained using prototype drift chambers (DC) with a 
construction similar to that anticipated for the final ALICE TRD 
\cite{aa:tdr}, but with a smaller active area (25$\times$32~cm$^2$).
To allow the measurement of the pure ionization energy loss for electrons,
no radiator was used for the present measurements.
The prototypes have a drift region of 30~mm and an amplification region 
of 7~mm.
Anode wires (W-Au) of 20~$\mu$m diameter are used, with a pitch of 5~mm.
The cathode wires (Cu-Be) have 75~$\mu$m diameter and a pitch of 2.5~mm.
We read out the signal on a segmented cathode plane with rectangular pads 
of 8~cm length and 0.75~cm width.
The entrance window (25~$\mu$m aluminized Kapton) simultaneously serves  
as gas barrier and as drift electrode.
We operate the drift chambers with the standard gas mixture for the TRD, 
Xe,CO$_2$(15\%), at atmospheric pressure.
The gas is recirculated using a dedicated gas system.
Our nominal gain of about 4000 is achieved with an anode voltage of 1.55~kV.
For our nominal drift field of 0.7 kV/cm, the detector signal is spread
over about 2~$\mu$s.

We use a prototype of the charge-sensitive preamplifier/shaper (PASA) 
especially designed and built for the TRD in 0.35~$\mu$m CMOS technology. 
It has a noise on-detector of about 1000 electrons r.m.s. and the FWHM 
of the output pulse is about 100~ns for an input step function.
The nominal gain of the PASA is 12~mV/fC, but during the present measurements 
we use a gain of 6~mV/fC for a better match to the range of the employed 
Flash ADC (FADC) system with 0.6~V voltage swing.
The FADC has adjustable baseline, an 8-bit non-linear conversion
and 20~MHz sampling frequency.
The FADC sampling was rebinned in the off-line analysis to obtain 100~ns 
time bins as for the final ALICE TRD \cite{aa:tdr}. 
The data acquisition (DAQ) was based on a VME event builder and was developed 
at GSI \cite{aa:mbs}.

\begin{figure}[htb]
\vspace{-.7cm}
\centering\includegraphics[width=.65\textwidth]{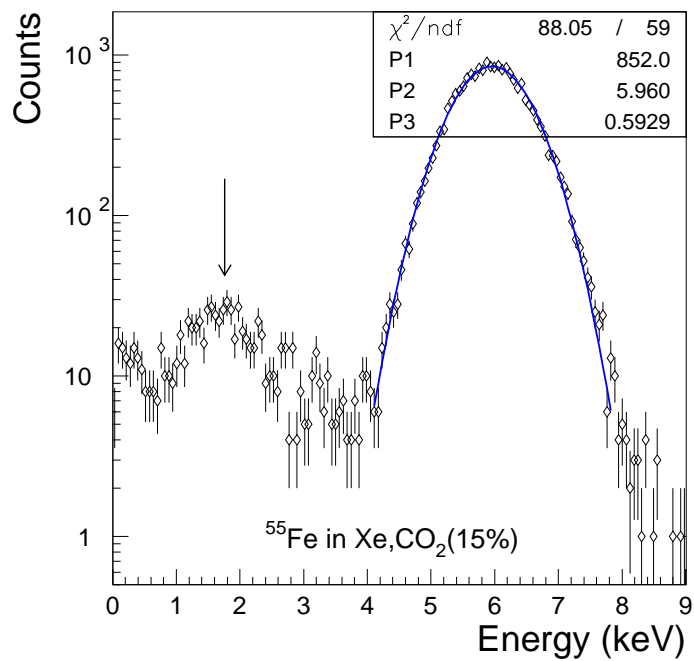}
\caption{$^{55}$Fe spectrum in Xe,CO$_2$(15\%).
A Gaussian fit of the main peak of 5.96~keV is also plotted. 
The arrow marks the expected position of the escape peak of Xe at 1.76~keV.}
\label{d:e1}
\end{figure}

The energy calibration of the detectors has been performed using a $^{55}$Fe
X-ray source. A spectrum of the integrated $^{55}$Fe signal is shown in
Fig.~\ref{d:e1} for our nominal anode voltage of 1.55~kV. The spectrum 
has been calibrated with one gain parameter (the baseline is obtained using 
empty presamples, there is no offset) using the main peak of 5.96~keV. 
The arrow marks the expected position of the escape peak of Xe at 1.76~keV, 
(the weighted average of the $L$ lines of Xe is 4.202 keV),
which is in good agreement with the measurements.
The energy resolution for the main peak is 23\% FWHM.

Four identical drift chambers were used for the beam measurements, 
without any radiator in front.
The variation of the gas gain for each individual chamber is within 10\%
and is calibrated away.
The measurements are carried out at momenta of 1, 1.5, 2, 3, 4, 5, 
and 6~GeV/c at the T10 secondary beamline of the CERN PS \cite{aa:cernpi}.
The momentum resolution is $\Delta p/p\simeq 1\%$.
The beam intensity is up to 3000 particles per spill of about 1 second.
As the beam diameter was of the order of 2.5~cm, we usually limited the 
readout of the DC to 8 pads. This also minimizes data transfer on the 
VSB bus connecting the FADC and the event builder.
The beam is a mixture of electrons and negative pions.
Similar sample sizes of pion and electron events are acquired within the
same run via dedicated triggers.
For the present analysis we have selected clean samples of pions and electrons
using coincident thresholds on two Cherenkov detectors and on a lead-glass 
calorimeter \cite{aa:andr}.
The incident angle of the beam with respect to the normal to the anode wires 
(drift direction)is 15$^\circ$ to avoid gas gain saturation due to space 
charge \cite{aa:gain}.

\section{Procedure and inputs for energy loss calculations} \label{d:sim}

The calculations are performed using a standalone Monte Carlo program 
especially written for this purpose.
In general, the ingredients needed for calculating the ionization energy 
loss of an incoming charged particle in any material are just 
the number of the primary inelastic collisions and 
the spectrum of the energy transfer in these collisions.
These quantities depend on the particle type and momentum (or Lorentz 
factor $\gamma$) and on the medium traversed.
In our case, the inputs were extracted from GEANT3 \cite{aa:geant}.

\begin{figure}[htb]
\vspace{-.5cm}
\centering\includegraphics[width=.67\textwidth]{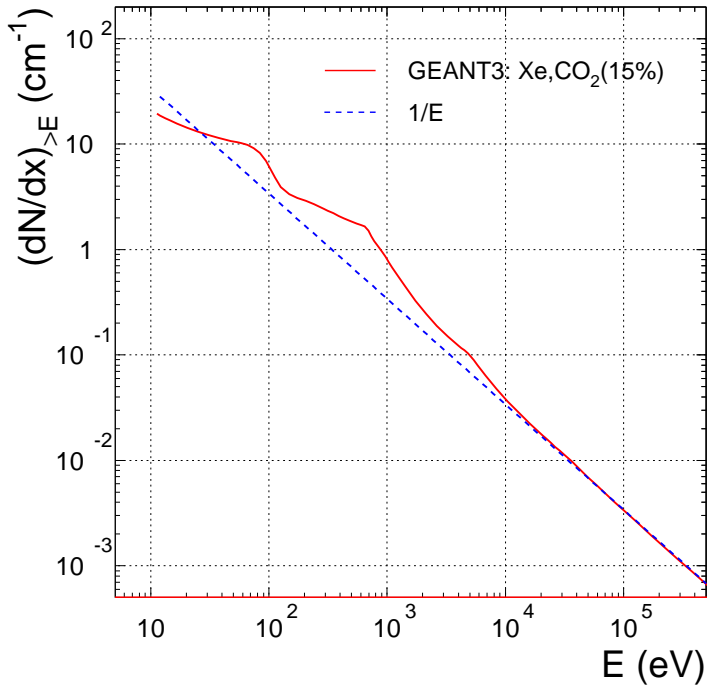}
\caption{The integral spectrum of the energy transfer in primary collisions.}
\label{d:sim1}
\end{figure}

The energy spectrum of primary electrons released in inelastic collisions 
of a minimum ionizing particle (MIP, $\gamma$=4) is presented in 
Fig.~\ref{d:sim1}. 
In GEANT3 an implementation of the photoabsorption ionization model \cite{aa:ali}
is used to calculate this spectrum. 
This is the integral spectrum of the number of the inelastic collisions 
per centimeter with an energy transfer greater than $E$,
$(\ud N/\ud x)_{>E}$.
For comparison, we plot also, arbitrarily normalized, the pure Rutherford 
spectrum, which, in the integral form, has a 1/$E$ dependence. 
The calculated spectrum has structures coresponding to the atomic shells
(most conspicous are the N and M shells of Xe) and approaches the Rutherford 
limit for large values of the energy transfer. 

The total number of primary collisions per cm of traversed gas, $N$, 
is the value corresponding to the lower limit of the energy transfer 
and is 19.3 for our case.
Note that this number is much smaller compared than the values quoted 
in earlier works: 48 by Ermilova et al. \cite{aa:erm} and 44 by Sauli 
\cite{aa:sauli} or Zarubin \cite{aa:zar}.
On the other hand, Va'vra \cite{aa:vav} has recently deduced that $N$
is about 25 for Xe.
The lowest value of the energy transfer in the GEANT3 spectrum is 11.26~eV
\cite{aa:geant}.
Since the value of the lowest ionization potential for Xe is $I$=12.13~eV 
(for CO$_2$ $I$=13.81~eV) \cite{aa:blum}, we restrict for our simulations  
the lowest energy transfer to this value of 12.13~eV, while keeping unchanged
the number of primary collisions.
This threshold leads to an increase of the simulated average energy loss 
by 4\% compared to the case when 11.26~eV would be the lowest allowed
energy transfer.

\begin{figure}[hbt]
\vspace{-.5cm}
\centering\includegraphics[width=.67\textwidth]{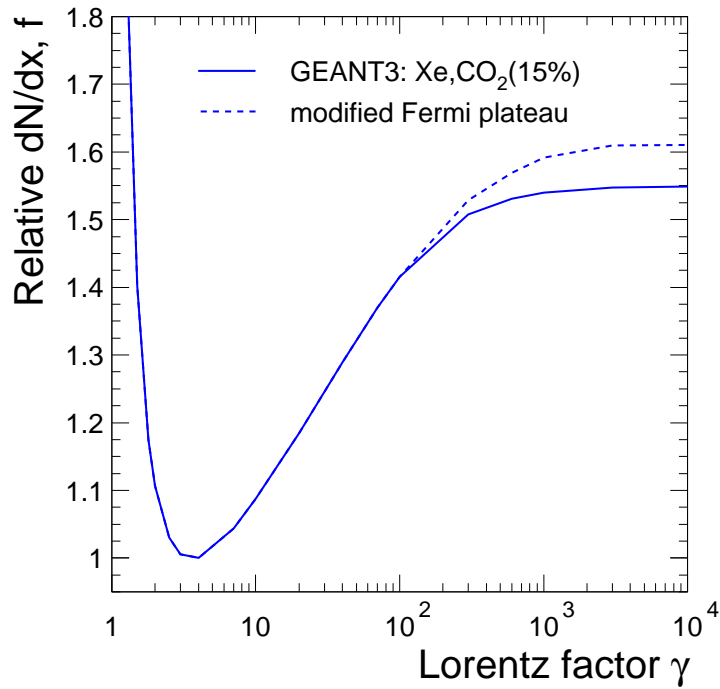}
\caption{The dependence of the primary number of inelastic collisions 
normalized to MIP ($\gamma$=4) on the Lorentz factor. The full line is
obtained using GEANT3, the dashed line is for an increased value
of the Fermi plateau that best reproduces our measurements.}
\label{d:sim2}
\end{figure}

In addition to the material-dependent values presented above, one has 
to specify their dependence on the incoming particle's Lorentz factor, 
$\gamma$. 
We assume that the spectral shape of the energy transfer is independent
of $\gamma$ and only the total number of primary collisions depends
on $\gamma$.
This factor, $f$, is the relative increase of $N$ with respect
to the value for MIP.
Its dependence, as extracted from GEANT3, is presented in Fig.~\ref{d:sim2} 
for our gas mixture.
Note that in GEANT3 the Fermi plateau corresponds to a number of primary 
inelastic collisions 1.55 times larger with respect to MIP. This value 
is sizeably larger than the value of 1.36 calculated by Ermilova et al. 
\cite{aa:erm}.
As we shall see below, our measurements favor an even higher Fermi plateau
than that extracted from GEANT3, of about 1.61. This is shown by the 
dashed line in Fig.~\ref{d:sim2}.
For a given detector thickness $D$ and for a given $\gamma$ value, 
the spectrum presented in Fig.~\ref{d:sim1} is sampled on average
$N_{tot}=NDf(\gamma)$ times. For an individual track, the actual 
number of inelastic collisions is obtained by sampling a Poissonian 
distribution with the average $N_{tot}$.

\begin{figure}[hbt]
\vspace{-.7cm}
\centering\includegraphics[width=.67\textwidth]{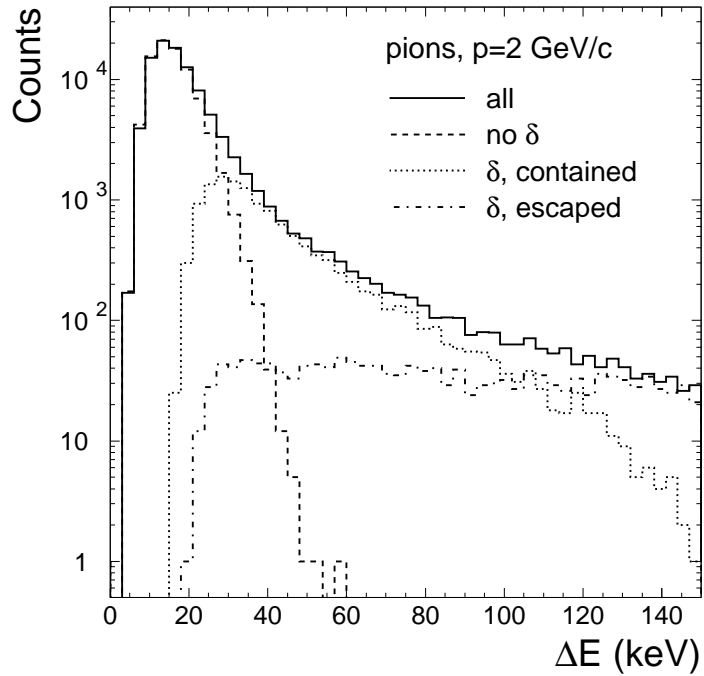}
\caption{Calculated integrated energy loss in our detector for 2 GeV/c 
pions for: i) all tracks (full histogram), ii) tracks which do not have any 
$\delta$-ray (dashed), iii) tracks with $\delta$-rays which are contained 
in the detector (dotted) and iv) tracks for which a $\delta$-ray has escaped 
the detector volume (dot-dashed).}
\label{d:sim3}
\end{figure}

The $\delta$-rays (energetic electrons produced in inelastic collisions) 
are tracked for energies above 10~keV.
This class is what we denote here generically $\delta$-rays (there is 
no accepted threshold above which a primary electron is called $\delta$-ray).
For these electrons the range is calculated according to the formula 
\cite{aa:blum}:
\begin{equation}
R(E)=AE\left( 1-\frac{B}{1+CE}\right),
\end{equation}
with $A$=5.37$\cdot$10$^{-4}$~gcm$^{-2}$keV$^{-1}$, $B$=0.9815, and 
$C$=3.123$\cdot$10$^{-3}$~keV$^{-1}$ \cite{aa:blum}.
$E$ is the energy of the $\delta$-ray.
For instance, for our gas mixture, the range is 0.52~mm for 10~keV and 
27.4~mm for 100 keV $\delta$-rays.
We assume that the $\delta$-rays move on a straight trajectory co-linear
with the parent particle.
If a $\delta$-ray was produced in the detector depth such that the range 
is greater than the remaining path length of the parent track in 
the detector, the $\gamma$ of the $\delta$-ray is calculated and its 
energy deposit is treated as for an independent track and added to 
the parent track.
For our detector geometry, 15.3\% of the tracks of 2 GeV/c pions have 
a $\delta$-ray above 10~keV. Of those, 16.3\% (or 2.5\% of all the tracks) 
escape from the detector volume.

In Fig.~\ref{d:sim3} we present the comparison of the energy loss 
integrated over the detector thickness, $\Delta$, for pions of 2~GeV/c.
The four cases are for: i) all tracks, ii) tracks not containing any 
$\delta$-ray, ii) tracks with $\delta$-ray, but which are completely absorbed 
in the detector, and iv) tracks for which one $\delta$-ray escapes the detector. 
For this last case the energy loss in this example is calculated taking 
the total energy of the $\delta$-ray.
These spectra illustrate the contribution of $\delta$-rays to the well
known Landau shape of energy loss in thin detectors: the tail originates
from the tracks for which one or more $\delta$-rays have been produced
\cite{aa:her}.
Depending on the detector thickness, some of those will escape the detection 
volume. This leads to the distinction between the energy loss and the 
energy deposit, which comprises only the detected signal (and is obviously
smaller than the energy loss).
The flat energy loss distribution for tracks containing escaped $\delta$-rays 
(dot-dashed histogram) is the result of the random distribution of these 
$\delta$-rays in the detector thickness: the lower energy $\delta$-rays 
escape only if they are produced at the end of the path of the parent particle
in the detector, while the very energetic ones escape no matter where they 
are produced.

\section{Results and discussion} \label{d:res}

\begin{figure}[htb]
\vspace{-.7cm}
\centering\includegraphics[width=.63\textwidth]{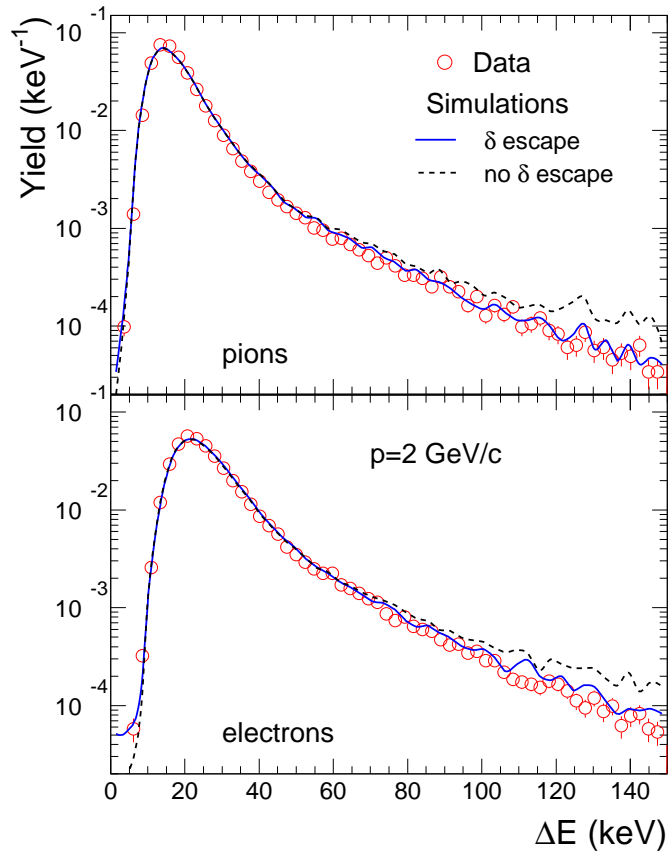}
\caption{Spectra of the energy loss of pions and electrons for the 
momentum of 2~GeV/c. The symbols represent the measurements. The lines 
are the simulations, with (continuous line) and without (dashed line) 
taking into account the finite range of $\delta$ electrons.}
\label{d:xe1}
\end{figure}

\begin{figure}[htb]
\vspace{-.7cm}
\centering\includegraphics[width=.52\textwidth]{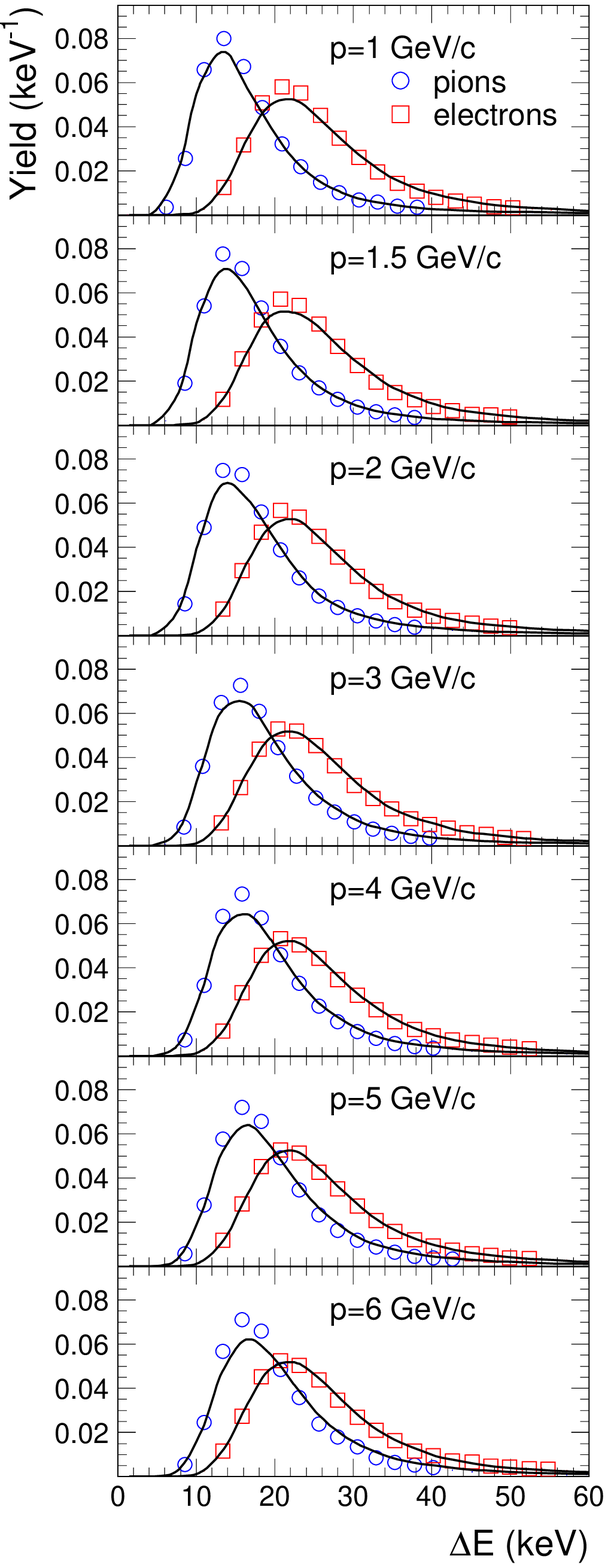}
\caption{The energy loss distributions of pions and electrons for momenta 
of 1 to 6~GeV/c. The symbols represent the measurements, the lines 
are the simulations.}
\label{d:xe1b}
\end{figure}

In Fig.~\ref{d:xe1} we present the measured distributions of energy loss
for pions and electrons for the momentum of 2~GeV/c.
Our high-statistics measurements with FADCs with large dynamic range 
make available for the first time complete energy loss spectra.
A very good agreement between data and calculations is seen for the case 
of $\delta$-rays escaping the detector volume. As noted above, those 
$\delta$-rays influence only the tails of the distributions.
This is the reason why the most probable value (MPV) of the spectrum 
and not its mean value is most commonly used to characterize the energy loss 
in thin detectors.
From here on we consider only the simulated results taking into account
the tracking of escaped $\delta$-rays.
The measurements for all momenta are presented in Fig.~\ref{d:xe1b}
together with simulations.
One can notice that, due to the relativistic rise of the pions, the separation 
between the pions and electrons is reduced as a function of momentum.
This is the reason why one employs the extra contribution of transition 
radiation in order to achieve a good electron/pion identification with
a TRD. 
In general, the agreement between data and calculations is good for all
momenta, but on a linear scale one can already notice that the widths 
of the distributions are larger in case of simulations (see below).
The calculated distributions are for the modified Fermi plateau.
All the measured spectra are available at the web page of our collaboration 
\cite{aa:trd}.
Note that, due to the track incidence at 15$^\circ$ with respect to the 
normal incidence, the distributions correspond to an effective detector 
thickness of 3.83~cm.

\begin{figure}[htb]
\vspace{-.9cm}
\centering\includegraphics[width=.65\textwidth]{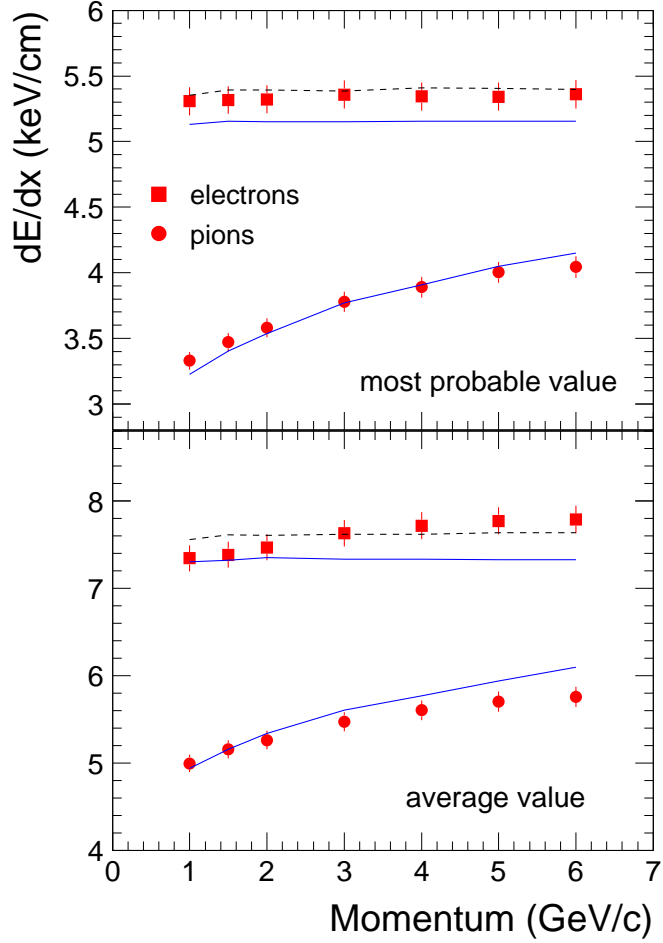}
\caption{Momentum dependence of the most probable and the average value of 
the energy loss for pions and electrons. The symbols are measurements, 
the lines are calculations (dashed line for the modified Fermi plateau).}
\label{d:xe2}
\end{figure}

In Fig.~\ref{d:xe2} we present the momentum dependence of the energy loss 
values for electrons and pions, for two cases: MPVs and average values. 
The MPVs are extracted from fits with a convolution of a Landau function 
and a Gaussian.
The measured data are compared to the results of simulations.
The errors on the data represent an estimated 2\% point-to-point accuracy, 
originating from minute changes of the gas composition and pressure.
We estimate an overall uncertainty of the absolute energy calibration 
of about 5\%. 
The statistical error of the Landau fit is negligible.
As noted earlier, for our momentum values, the pions are in the regime 
of the relativistic rise, while the electrons are at the Fermi plateau.
These two regimes are reflected in the measured values of the energy loss.
Note that our average values for electrons are lower than those reported 
by Fischer et al. \cite{aa:fis}, who measured a value of about 
9~keV/cm at the Fermi plateau.
Appuhn et al. \cite{aa:zeus} have reported for electrons even larger values 
of about 12~keV/cm, with a slightly increasing trend for the same momentum
range as ours.

The calculations reproduce well the absolute magnitude and the general trends 
of the data, however, in case of electrons, only with the modified value of
the Fermi plateau.
The calculations show a more pronounced relativistic rise than the measurements.
Also, in case of the average values, the calculations indicate that the Fermi 
plateau is reached for the electrons for all momentum values, while the data 
show a slight increase for momenta below 2 GeV/c.
Despite the above-mentioned 5\% overall uncertainty of the measured values,
it is evident that the calculations cannot consistently explain the measured 
values for pions and electrons unless one introduces the modified Fermi
plateau.

\begin{figure}[htb]
\vspace{-.7cm}
\centering\includegraphics[width=.67\textwidth]{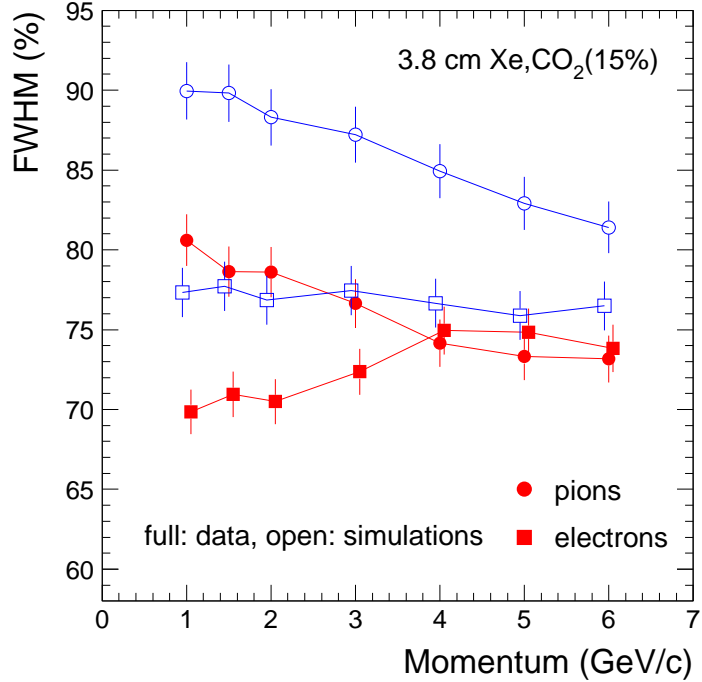}
\caption{Momentum dependence of the FWHM value of the energy loss normalized 
to MPV for pions and electrons. Full symbols are measurements, open symbols 
are calculations.}
\label{d:xe3}
\end{figure}

The comparion of the measured and simulated values of the FWHM of the
energy loss spectra, normalized to MPV, is shown in Fig.~\ref{d:xe3}
for pions and electrons as a function of particle momentum. 
The FWHM is determined by the magnitude of the ionization, in particular
by the average number of primary inelastic collisions. 
For pions, as a consequence of the relativistic rise, one expects that 
the FWHM is decreasing as a function of momentum and this is indeed the 
trend of our measured values. 
For electrons, due to their constant energy loss, the FWHM is expected 
to be constant, but the measurements show a slight increase as a function 
of momentum. This is not quantitatively understood, but Bremsstrahlung 
is the only candidate to explain such a behavior.
Our measured values for pions are larger than those reported in 
ref.~\cite{aa:onu} by about 15\%.

The calculations reproduce the measured trend for pions, but show clearly 
larger values of FWHM.
The expected constant value of FWHM for electrons is confirmed by the 
calculations. Bremsstrahlung is not included for the simulated events.
Larger FWHMs in case of calculations may be a consequence of too low 
a number of primary inelastic collisions predicted by GEANT. 
Obviously, one can not increase this number unconditionally.
The spectrum of the energy transfer has to be changed in this case too,
otherwise the average values of dE/dx would no longer be in agreement with 
the measurements.
As emphasized by the authors of GEANT3 \cite{aa:geant}, the shape of
the energy transfer spectrum at low energy is strongly dependent on the 
choice and treatment of the photo-absorption cross sections.
If the FWHM is only determined by the number of primary collisions, a simple
Poisson scaling implies that about 25 primary collisions per cm would be
required for an agreement between our measurements and the simulations
in case of pions.
This value is identical to the one inferred by Va'vra \cite{aa:vav}.
Another possibility is that for the measurements the Penning effect 
increases the secondary electron statistics \cite{aa:zar,aa:bia}.
We note that similarly larger widths of simulated distributions compared 
to measurements are also apparent in ref.~\cite{aa:apo}.

\begin{figure}[htb]
\vspace{-.7cm}
\centering\includegraphics[width=.7\textwidth]{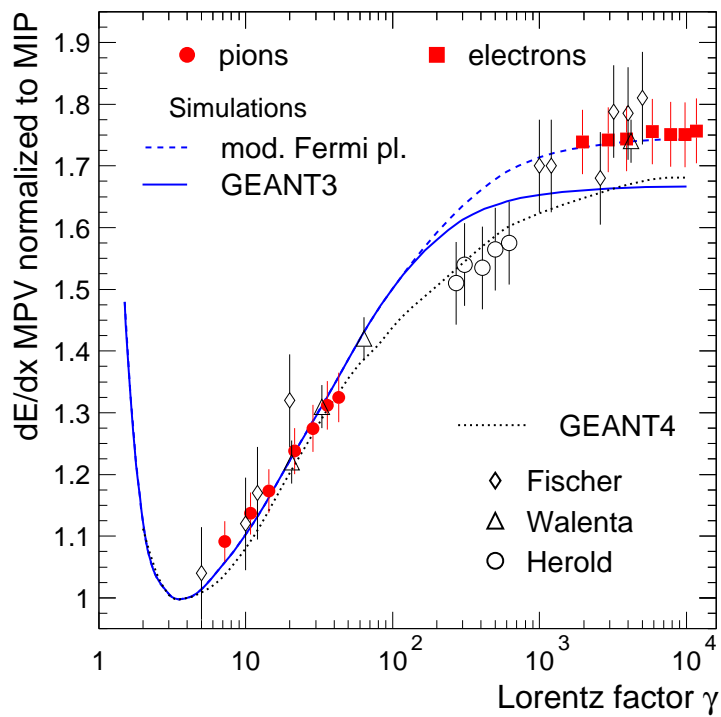}
\caption{Relativistic rise of the most probable value of the energy loss.
Our values (full symbols) are compared to calculations and to the
measurements of Fischer et al. \cite{aa:fis}, Walenta et al. \cite{aa:wal} 
and Herold et al. \cite{aa:her}. The dashed line is the result of GEANT4 
calculations \cite{aa:apo}.}
\label{d:xe4}
\end{figure}

In Fig.~\ref{d:xe4} we present the relativistic rise of the most probable 
value of the energy loss. Our data are compared to other measurements 
\cite{aa:fis,aa:wal,aa:her} and to calculations. 
For this relative comparison, we have normalized our measured value for 
3~GeV/c pions ($\gamma$=21) to the calculations. Note that the absolute 
MPVs of data and calculations agree perfectly at this momentum value 
(see Fig.~\ref{d:xe2}).
The agreement with existing measurements is very good. 
In particular, we would like to emphasize the perfect agreement with
the Fermi plateau value of Walenta et al. \cite{aa:wal}.
As expected from the previous comparisons, the calculations reproduce the 
data well, but only in case of the modified Fermi plateau.
The results of GEANT4 calculations \cite{aa:apo} are included as well.
The value of the Fermi plateau in GEANT4 is very close to the GEANT3 case 
and, as a consequence, underpredicts our measurements.
The approach to the plateau is more gradual in case of GEANT4 and
this agrees with the measurements of Herold et al. \cite{aa:her}, which
are clearly overpredicted by our calculations.
A Fermi plateau of about 1.98 was calculated \cite{aa:fis} using the
density effect correction of Sternheimer and Peierls \cite{aa:ste}.
On the other hand, Cobb et al. \cite{aa:cob} have calculated a value of
1.73, very close to our measured value.
Note that the value of the Fermi plateau in the case of MPV, 1.75, 
is larger compared to the corresponding value for the average energy loss, 
1.61 (which is obviously identical to the input value for the Fermi plateau
of the number of primary collisions), and the onset of the plateau starts 
at somewhat larger values of $\gamma$ (see also Fig.~\ref{d:sim2}) \cite{aa:lap}.

\begin{figure}[htb]
\vspace{-.7cm}
\centering\includegraphics[width=.68\textwidth]{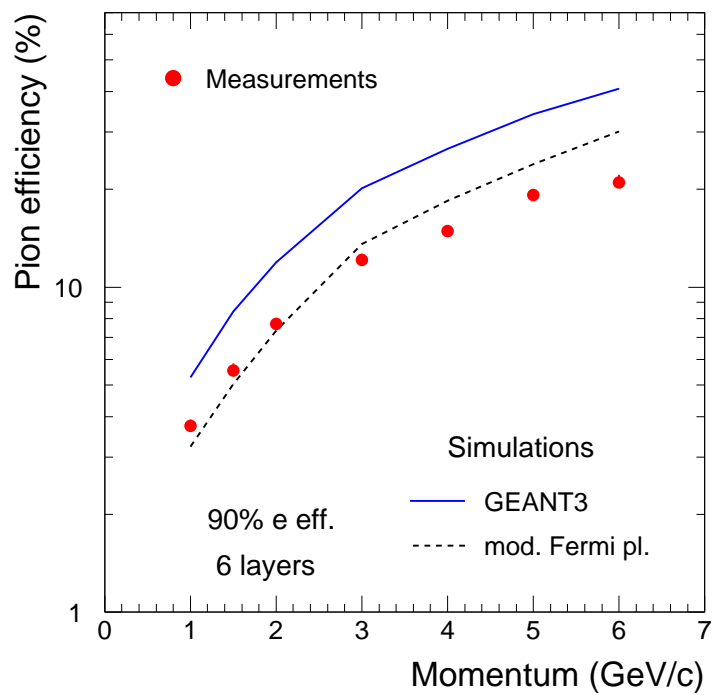}
\caption{The momentum dependence of the pion efficiency for 90\% electron
efficiency. The measurements for 4 layers are used to calculate the
expected performance for 6 layers. The lines are the results of simulations.}
\label{d:xe5}
\end{figure}

The results presented above allow the calculation of the electron/pion 
rejection using pure dE/dx measurements. 
The results are presented in Fig.~\ref{d:xe5}. 
A likelihood method \cite{aa:bun} on the total energy loss spectra 
(see Fig.~\ref{d:xe2}) has been used to extract the pion efficiencies 
for 90\% electron efficiency.
The measured experimental data for 4 layers has been used to calculate the 
expected performance for the final ALICE TRD configuration with 6 layers. 
The experimental errors are of the order of the dimension of the points.
The data are compared to the performance extracted from the simulations
using an identical procedure. Clearly, the modified Fermi plateau is needed
in order to explain the measured pion efficiencies. Moreover, as a result
of the faster relativistic rise in the simulations compared to the data,
the simulations show a degradation of the pion rejection performance
which is steeper than the measurements. 
The larger widths of the simulated Landau distributions play a role, too.

In ALICE TRD the contribution of TR is significantly improving the electron 
identification performance \cite{aa:andr}. 
Pion efficiencies below 1\% have been achieved in tests with prototypes
\cite{aa:andr}.
However, particle identification using dE/dx will be used for other charged 
particles, for which truncated means could be a more advantageous
way to exploit the measured signal.
In case of electron/pion separation, the comparison of measurements and 
calculations in terms of pure dE/dx is just a necessary step in understanding 
the contribution of TR, in particular the comparison between measurements 
and simulations of the TRD performance.

\section{Summary} \label{aa:sum}

We have reported measurements of ionization energy loss in drift chambers
operated with Xe,CO$_2$(15\%), carried out using prototype drift chambers 
for the ALICE TRD.
Pions and electrons with momenta from 1 to 6~GeV/c were studied.
Our high-statistics measurements with FADCs with large dynamic range 
make available for the first time complete energy loss spectra.
Our measured relativistic rise agrees well with existing measurements.
The measured distributions are in general well reproduced using 
microscopic calculations with GEANT3 input parameters, but a modified value 
of the Fermi plateau (an increase from 1.55 to 1.61) is needed to explain 
the electron data.
The calculations show wider distributions compared to the measurements,
suggesting that the number of primary collisions is too low in the standard
GEANT3 input values. 
Apparently, all the noticed discrepancies between data and GEANT3 will apply
in case of GEANT4, too.

\section*{Acknowledgments}
We acknowledge A. Radu and J. Hehner for the skills and dedication in building
our detectors.
We acknowledge useful discussions with S. Biagi.
We would also like to acknowledge P. Szymanski for help in organizing the
experiment and A. Przybyla, and M. Wensveen for technical assistance during 
the measurements. We thank N. Kurz for assistance with the data acquisition.


\end{document}